\documentstyle[12pt]{article}
\renewcommand{\thefootnote}{\fnsymbol{footnote}}

\newcommand{\EQ}{\begin{equation}}
\newcommand{\EN}{\end{equation}}
\newcommand{\bea}{\begin{eqnarray}}
\newcommand{\ena}{\end{eqnarray}}
\newcommand{\vs}[1]{\vspace{#1 mm}}

\newcommand{\uda}{\nearrow \kern-1em \searrow}

\newcommand{\CMP}[1]{Comm.\ Math.\ Phys.\ {\bf #1}}

\newcommand{\PTP}[1]{Prog.\ Theor.\ Phys.\ {\bf #1}}

\newcommand{\AJ}[1]{Astorophys. \ J.\ {\bf #1}}
\newcommand{\JMP}[1]{J.\ Math.\ Phys.\ {\bf #1}}

\begin{document}
\setlength{\baselineskip}{7mm}

\begin{titlepage}
\setcounter{page}{0}
\begin{flushright}
EPHOU 98-005\\
OU-HET-296\\
May 1998
\end{flushright}
\vs{6}
\begin{center}
{\Large Perturbations of Kerr-de Sitter Black Hole and Heun's Equations}

\vs{5}
{\bf {\sc
Hisao Suzuki\footnote{e-mail address: hsuzuki@particle.sci.hokudai.ac.jp}}
\\
{\em Department of Physics, \\
Hokkaido
University \\  Sapporo, Hokkaido 060 Japan} \\
{\sc Eiichi Takasugi\footnote{
e-mail address: takasugi@phys.wani.osaka-u.ac.jp}}
\\and \\
{\sc Hiroshi Umetsu\footnote{
e-mail address: umetsu@phys.wani.osaka-u.ac.jp}}\\
{\em Department of Physics,\\
Osaka University\\
Toyonaka, Osaka 560 Japan}}\\
\end{center}
\vs{6}

\begin{abstract}
It is well known that the perturbation equations of massless fields for 
the Kerr-de Sitter geometry can be written in the form of 
separable equations. The equations  have five definite singularities so that 
the analysis has been expected to be difficult. We show that these equations 
can be transformed to  Heun's equations, for which we are able to use 
known technique for the analysis of the solutions.  
We reproduce results known previously for the Kerr geometry and 
de Sitter geometry in the confluent limits of the Heun's functions. 
Our analysis applies can be extended to Kerr-Newman-de Sitter geometry for 
massless fields with spin 0 and $\frac 12$.  
\end{abstract}
\end{titlepage}

\renewcommand{\thefootnote}{\arabic{footnote}}
\setcounter{footnote}{0}


\section{Introduction}
One of the most non-trivial aspects of the perturbation 
equations  for Kerr geometry is 
the separability of the radial and the angular parts. 
It was Carter[1] who first found that the scalar 
wave function is separable in the Kerr-Newman-de Sitter geometries. 
Later, this observation has been extended for spin 1/2, 
electromagnetic fields, gravitational perturbations and gravitino 
for the Kerr geometries and even for the Kerr-de Sitter class of 
geometries. The equations are called Teukolsky equations[2]. 
Except for the electromagnetic and gravitational perturbations, 
the separability persists even for the Kerr-Newman-de Sitter solutions. 
An important application of this fact was the proof of the 
stability of the Kerr black hole[3]. 

Though the Teukolsky equation for Kerr geometries is separable, 
both angular and radial equations has two regular and 
one irregular singularities so that the solutions cannot be written by 
a single form of any special functions, 
but they are expressed as series of special 
functions whose coefficients satisfy the three term recurrence 
relations[4-7]. 
The solution of the angular equation is expressed in the form of 
series of Jacobi functions[4]. The solution of the radial equation is rather 
complicated because we need the solution which is valid in the entire 
region extending from the 
outer horizon to infinity. The solution is written in the form of series of 
confluent hypergeometric functions[5] which is convergent around infinity 
and the hypergeometric functions[6-7] which is convergent around the 
outer horizon. By matching these solutions in the region where both solutions 
are convergent, we can obtain the solution which is valid from the 
outer horizon to infinity[6-8]. The great benifit 
of this kind of solutions is that the coefficients of the series are 
obtained as the Post Minkowskian expansion[6-7]. This technique has been 
successfully applied for the Post Newtonian expansion of 
gravitational waves from a particle in circular orbits around a rotating 
black hole[9-10]. 

Strictly speaking, the technique of constructing solutions as 
series of special functions is not new, but extremely old. The similar 
expansion has been considered in the case of Heun's equation[11-12]. 
It may  not be difficult to see the 
correspondense between  the Teukolsky equations for 
the Kerr black hole and the Heun's equation. That is, the Teukolsky 
equation with two regular and  one irregular singularities will  be  
given as a confluent case of the Heun's equation which has 
four definite singularities as we can see from Eq.(2.2) in Ref.6 
and page 27 of Ref.12. 
 
It is  known that the separability holds true for 
the Kerr-de Sitter geometries and even for the Kerr-Newman-de Sitter 
geometries except for electormagnetic and gravitational fields. 
Therefore, it may be meaningful to 
consider whether the Teukolsky equations for these 
geometries can be transformed into  the Heun's equations, although 
these Teukolsky equations have five regular singularities. 
A usuful suggestion for this problem can be seen in  the 
radial part of Teukolsky equation for the de Sitter geometries[13].  
We showed that although this equation has four regular singularities,  
the one of the singularities can be eliminated by a suitable change 
of variables and by the redefinition of the radial function, so that 
the solutions of the equation are given by the hypergeometric function. 
From this experience, we expect that the similar non-trivial 
transformations may exist  even for the Kerr(-Newman)-de Sitter 
geometries.

Our main aim of this short letter is to show that the Teukolsky 
equations for Kerr-Newman-de Sitter geometries 
can be transformed into the Heun's equations. Here, we stress again 
that our discussions for the Kerr-de Sitter geometries 
appliy to massless particles, but those for the Kerr-Newman-de Sitter 
geometries do apply to massless particles except for electromagnetic 
and gravitational fields. 
Then, we can use very old technique[12] 
for the analysis of the solutions. 
The previous solutions for the Kerr geometries[6-7] and 
de Sitter geometries[13] can be obtained by confluent limits of 
the Heun's functions. 

In section 2, we provide a short review of 
the Teukolsky equations for the Kerr-Newman-de Sitter geometires. 
In section 3, we show that both angular and radial 
equations can be transformed into the Heun's equations. 
In section 4, we analize the solutions of the angular and radial equations 
by using the results of the Heun's equation and show that the previously 
obtained solutions for the Kerr geometries and de Sitter geometries 
can be obtained  by the confluent limits of the soluitions for 
Kerr(-Newman)-de Sitter geometries. 
The last section is devoted to some discussions and further possible 
applications.


\section{Teukolsky equation for the Kerr(-Newman)-de Sitter geometry}
\setcounter{equation}{0}

We consider the Teukolsky equations for the Kerr-Newman-de Sitter 
geometries. Since the electromagnetic field and the gravitational 
field couple through electric charge of black hole in this geometry 
and the equation is not separable so far, we deal with massless 
fields with any spins except for these fields. 
In the limit of null electric charge, $Q \to 0$, 
the Kerr-de Sitter geometries are realized where the Teukolsky 
equation is valid for all massless fields.  
In the Boyer-Lindquist coordinates the Kerr-Newman-de Sitter metric has 
the form,
\begin{eqnarray}
ds^2 &=& -\rho^2
\left(\frac{dr^2}{\Delta_r}+\frac{d\theta^2}{\Delta_\theta}\right)
-\frac{\Delta_\theta \sin^2\theta}{(1+\alpha)^2 \rho^2}
[adt-(r^2+a^2)d\varphi]^2 \nonumber \\ 
&&  +\frac{\Delta_r}{(1+\alpha)^2 \rho^2}(dt-a\sin^2\theta d\varphi)^2,
\end{eqnarray}
where 
\begin{equation}
\begin{array}{cc}
\multicolumn{2}{c}{{\displaystyle 
\Delta_r=(r^2+a^2)\left(1-\frac{\alpha}{a^2}r^2\right)-2Mr+Q^2
=-\frac{\alpha}{a^2}(r-r_+)(r-r_-)(r-r'_+)(r-r'_-),}} \\
\Delta_\theta=1+\alpha\cos^2\theta, &
{\displaystyle \alpha=\frac{\Lambda a^2}{3}}, \\
\bar{\rho}=r+ia\cos\theta, &
\rho^2=\bar{\rho}\bar{\rho}^*. \\
\end{array}
\end{equation}
Here $\Lambda$ is the cosmological constant, $M$ is the mass of the black hole, $aM$ its angular momentum and $Q$ its charge.
The electromagnetic field due to the charge of the black hole 
is given by
\begin{equation}
A_{\mu}dx^{\mu}
   =-\frac{Qr}{(1+\alpha)^2\rho^2}(dt-a\sin^2\theta d\varphi).
\end{equation}
In particular we adopt the following vectors as the null tetrad,
\begin{eqnarray}
l^\mu &=& \left(\frac{(1+\alpha)(r^2+a^2)}{\Delta_r}, 1, 0, 
\frac{a(1+\alpha)}{\Delta_r}\right), \nonumber \\
n^\mu &=& \frac{1}{2\rho^2}\left((1+\alpha)(r^2+a^2), -\Delta_r, 0, 
a(1+\alpha)\right), \\
m^\mu &=& \frac{1}{\bar{\rho}\sqrt{2\Delta_\theta}}
\left(ia(1+\alpha)\sin\theta, 0, \Delta_\theta, 
\frac{i(1+\alpha)}{\sin\theta}\right), 
\makebox[1cm]{} \bar{m}^\mu={m^*}^\mu. \nonumber
\end{eqnarray}
Assuming that the time and azimuthal dependence of the fields 
has the form $e^{-i(\omega t-m\varphi)}$, the tetrad components of 
the derivative and the electromagnetic field are
\begin{equation}
\begin{array}{cc}
l^\mu \partial_\mu={\cal D}_0, & 
n^\mu \partial_\mu=\displaystyle{
-\frac{\Delta_r}{2\rho^2}{\cal D}_0^\dag}, \\
m^\mu \partial_\mu=\displaystyle{
\frac{\sqrt{\Delta_\theta}}{\sqrt{2}\bar{\rho}}
{\cal L}_0^\dag}, &
\bar{m}^\mu \partial_\mu=\displaystyle{
\frac{\sqrt{\Delta_\theta}}{\sqrt{2}\bar{\rho}^*}{\cal L}_0},\\
l^\mu A_\mu=\displaystyle{-\frac{Qr}{\Delta_r}}, &
n^\mu A_\mu=\displaystyle{-\frac{Qr}{2\rho^2}}, \\
m^\mu A_\mu=\bar{m}^\mu A_\mu=0, &
\end{array}
\end{equation}
where 
\begin{eqnarray}
{\cal D}_n&=&\partial_r-\frac{i(1+\alpha)K}{\Delta_r}
+n\frac{\partial_r \Delta_r}{\Delta_r}, \nonumber \\
{\cal D}_n^\dag&=&\partial_r+\frac{i(1+\alpha)K}{\Delta_r}
+n\frac{\partial_r \Delta_r}{\Delta_r},  \\
{\cal L}_n&=&\partial_\theta+\frac{(1+\alpha)H}{\Delta_\theta}
+n\frac{\partial_\theta(\sqrt{\Delta_\theta} \sin\theta)}
{\sqrt{\Delta_\theta} \sin\theta}, \nonumber \\
{\cal L}_n^\dag&=&\partial_\theta-\frac{(1+\alpha)H}{\Delta_\theta}
+n\frac{\partial_\theta(\sqrt{\Delta_\theta} \sin\theta)}
{\sqrt{\Delta_\theta} \sin\theta}, \nonumber
\end{eqnarray}
with $K=\omega(r^2+a^2)-am$ and
$\displaystyle{H=-a\omega\sin\theta+\frac{m}{\sin\theta}}$.

Using the Newman-Penrose formalism it is known that perturbation equations 
in the Kerr-de Sitter geometry are separable for massless 
spin 0, $\frac{1}{2}$, 1, $\frac{3}{2}$ and 2 fields.
Similarly in the Kerr-Newman-de Sitter space those for spin 0,
$\frac{1}{2}$ fields are also separable.
The separated equations for fields with spin $s$ and charge $e$ are 
given by 
\begin{eqnarray}
&& \left[
\sqrt{\Delta_\theta}{\cal L}_{1-s}^{\dag}\sqrt{\Delta_\theta}
{\cal L}_s
- 2(1+\alpha)(2s-1)a\omega\cos\theta \right.\nonumber \\
&&  \makebox[50mm]{} \left.
- 2\alpha(s-1)(2s-1)\cos^2\theta+\lambda \right]S_s(\theta) = 0, 
\label{eqn:Ss}\\
&& \left[
\Delta_r {\cal D}_1 {\cal D}_s^\dag +2(1+\alpha)(2s-1)i\omega r 
-\frac{2\alpha}{a^2}(s-1)(2s-1) \right.  \nonumber \\
&& \makebox[5mm]{}  \left.  
+\frac{-2(1+\alpha)eQKr+iseQr\partial_r \Delta_r+e^2Q^2r^2}
{\Delta_r}
- 2iseQ-\lambda \right]R_s(r) = 0. \label{eqn:Rs}
\end{eqnarray}
%


\section{Transformation of Teukolsky equation to Heun's equation}
\setcounter{equation}{0}
In this section we show that the Teukolsky equations (\ref{eqn:Ss}) 
and (\ref{eqn:Rs}) can be transformed to the Heun's equations by 
factoring out a single regular singularity.
\subsection{Angular Teukolsky equation} 
From eq.(\ref{eqn:Ss}), the angular Teukolsky equation become 
\begin{eqnarray}
\Bigg\{&& \makebox[-0.7cm]{} \frac{d}{dx} ( 1+\alpha x^2 ) (1-x^2) 
\frac{d}{dx}
+ \lambda - s (1-\alpha) + \frac{(1+\alpha)^2}{\alpha} \xi^2
- 2 \alpha x^2 \nonumber \\  
&+& \frac{1+\alpha}{1+\alpha x^2} 
\left[ \ 2s (\alpha m - (1+\alpha) \xi) x 
- \frac{(1+\alpha)^2}{\alpha} \xi^2 
-2 m (1+\alpha) \xi + s^2 \ \right] \nonumber \\ 
&-& \frac{(1+\alpha)^2 m^2}{(1+\alpha x^2) (1-x^2)}
- \frac{(1+\alpha) (s^2 + 2smx)}{1-x^2} \ \ \Bigg\} S(x) = 0, \label{eqn:sx}
\end{eqnarray}
where $x = \cos \theta$ and $\xi = a \omega$.
This equation has five regular singularities at $\pm 1$, 
$\pm \frac{i}{\sqrt{\alpha}}$ and $\infty$.
We also note that the angular equation has no dependence on $M$ and $Q$. 
By choosing the variable $z$
$$
z = \frac{1-\frac{i}{\sqrt{\alpha}}}{2} \frac{x+1}{x-\frac{i}{\sqrt{\alpha}}},
$$
then eq.(\ref{eqn:sx}) takes the following form,
\begin{eqnarray}
\Bigg\{ && \lefteqn{ \makebox[-0.7cm]{} \frac{d^2}{dz^2} 
+ \left[ \ \frac{1}{z} + \frac{1}{z-1} + \frac{1}{z-z_s}
- \frac{2}{z-z_\infty} \ \right] \frac{d}{dz} } \nonumber \\ 
&& \makebox[-0.7cm]{}
-\left(\frac{m-s}{2}\right)^2 \frac{1}{z^2} 
-\left(\frac{m+s}{2}\right)^2 \frac{1}{(z-1)^2} 
+\left(\frac{1+\alpha}{2\sqrt{\alpha}} \xi 
- \frac{\sqrt{\alpha}m+is}{2}\right)^2 
\frac{1}{(z-z_s)^2} 
+ \frac{2}{(z-z_\infty)^2} \nonumber \\
&& \makebox[-0.9cm]{}
+\left[-\frac{m^2}{2}\left(1+\frac{4\alpha}{(1+i\sqrt{\alpha})^2}\right)
+ \frac{s^2}{2}\left(\frac{1-i\sqrt{\alpha}}{1+i\sqrt{\alpha}}
\right)^2 
+\frac{2ims\sqrt{\alpha}}{1+i\sqrt{\alpha}} \right.\nonumber \\
&& \left.\hspace{3cm}
+\frac{\lambda-s(1-\alpha)-2\alpha+2(1+\alpha)(m+s)\xi}
{(1+i\sqrt{\alpha})^2} \ 
\right] \frac{1}{z} \nonumber \\
&& \makebox[-0.9cm]{}
+ \left[\frac{m^2}{2}\left(1+\frac{4\alpha}{(1-i\sqrt{\alpha})^2}\right)
- \frac{s^2}{2}\left(\frac{1+i\sqrt{\alpha}}{1-i\sqrt{\alpha}}
\right)^2
-\frac{2ims\sqrt{\alpha}}{1-i\sqrt{\alpha}} \right.\nonumber \\ 
&& \left.\hspace{3cm} 
-\frac{\lambda-s(1-\alpha)-2\alpha+2(1+\alpha)(m-s)\xi}
{(1-i\sqrt{\alpha})^2} \ 
\right] \frac{1}{z-1} \nonumber \\
&& \makebox[-0.9cm]{}
+\left[-m^2\frac{2i\alpha\sqrt{\alpha}}{(1+\alpha)^2}
- s^2\frac{i\sqrt{\alpha}(1-\alpha)}{(1+\alpha)^2}
- ms\frac{\alpha}{1+\alpha} 
+\frac{i\sqrt{\alpha}(\lambda-s(1-\alpha)+2)}{(1+\alpha)^2}
\right.\nonumber \\ 
&& \left.\hspace{2cm}
+\frac{(2i\sqrt{\alpha}m+(\alpha-1)s)\xi}{1+\alpha} \ 
\right] \frac{4}{z-z_s} 
-\frac{8i\sqrt{\alpha}}{1+\alpha} \frac{1}{z-z_\infty} \ 
\Bigg\} S(z) = 0,
\end{eqnarray}
where $\displaystyle{z_s=-\frac{i(1+i\sqrt{\alpha})^2}{4\sqrt{\alpha}}}$ 
and $\displaystyle{z_\infty=-\frac{i(1+i\sqrt{\alpha})}{2\sqrt{\alpha}}}$.
It turns out that the singularity at $z=z_\infty$ which  corresponds to 
$x=\infty$ can be factored out by the transformation,
\begin{equation}
S(z)=z^{A_1} (z-1)^{A_2} (z-z_s)^{A_3} (z-z_\infty) f(z),\label{eqn:Sf}
\end{equation}
where $\displaystyle{A_1=\frac{|m-s|}{2}}$, 
$\displaystyle{A_2=\frac{|m+s|}{2}}$ and 
$\displaystyle{A_3=\pm\frac{i}{2}\left(\frac{1+\alpha}{\sqrt{\alpha}}\xi
-\sqrt{\alpha}m-is\right)}$.
Now $f(z)$ satisfies the equation
\begin{equation}
\left\{ \ \frac{d^2}{dz^2} + \left[\frac{2A_1+1}{z} + \frac{2A_2+1}{z-1}
+\frac{2A_3+1}{z-z_s} \right] \frac{d}{dz}
+ \frac{\rho_+ \rho_- z + u}{z (z-1)(z-z_s)} \ \right\} f(z) =0,
\label{eqn:HS}
\end{equation}
where 
\begin{eqnarray}
\rho_{\pm}\makebox[-3mm]{}&=&\makebox[-3mm]{}
A_1+A_2+A_3 \pm A_3^* +1, \\
u\makebox[-3mm]{}&=&\makebox[-3mm]{}
\frac{-i}{4\sqrt{\alpha}} \Bigg\{
\lambda-s(1-\alpha)-2i\sqrt{\alpha}+2(1+\alpha)(m+s)\xi
-(1+i\sqrt{\alpha})^2 (2A_1 A_2+A_1+A_2) \nonumber \\
&&\makebox[1cm]{}
- 4i\sqrt{\alpha}(2A_1 A_3+A_1+A_3)
-\frac{m^2}{2}\left[4\alpha+(1+i\sqrt{\alpha})^2\right] \nonumber \\
&&\makebox[1cm]{} +\frac{s^2}{2}(1-i\sqrt{\alpha})^2
+2ims\sqrt{\alpha}(1+i\sqrt{\alpha}) \Bigg\}.
\end{eqnarray}
Equation (3.4) is called the Heun's equation 
which has four regular singularities. The $f(z)$ is determined by 
requiring non-singular behaviors at $z=0$ and 1. We can take 
either one of signs of $A_3$ to find the solution $S(z)$. 


\subsection{Radial Teukolsky equation}
From eq.(\ref{eqn:Rs}), the radial Teukolsky equation is explicitly 
written by 
\begin{eqnarray}
&&\makebox[-10mm]{}\Bigg\{ \ 
\Delta_r^{-s}\frac{d}{dr}\Delta_r^{s+1}\frac{d}{dr}
+\frac{1}{\Delta_r}\left[ (1+\alpha)^2 \left(K-\frac{eQr}{1+\alpha}
\right)^2
- is(1+\alpha)\left(K-\frac{eQr}{1+\alpha}\right) \frac{d\Delta_r}
{dr} \right] \nonumber \\
&& \makebox[-5mm]{}
+4is(1+\alpha)\omega r -\frac{2\alpha}{a^2}(s+1)(2s+1) r^2
+2s(1-\alpha)-2iseQ -\lambda \ \Bigg\} R = 0. \label{eqn:Rr}
\end{eqnarray}
This equation has five regular singularities at  $r_\pm, r'_\pm$ and 
$\infty$ which are assigned such that  
$r_\pm \rightarrow M\pm\sqrt{M^2-a^2-Q^2} \equiv r^0_\pm$ and 
$\displaystyle{r'_\pm \rightarrow \pm\frac{a}{\sqrt{\alpha}}}$ 
in the limit $\alpha \rightarrow 0$ $(\Lambda \rightarrow 0)$. 
By using the new variable 
$$
z=\left (\frac{r_+ - r'_-}{r_+ - r_-}\right )
\left ( \frac{r - r_-}{r - r'_-}\right ),
$$
eq (\ref{eqn:Rr}) becomes an equation which has regular singularities at 
$0, 1, z_r, z_\infty$ and $\infty$,
$$
z_r=\left(\frac{r_+ - r'_-}{r_+ - r_-}\right )\left(
 \frac{r'_+ - r_-}{r'_+ - r'_-}\right)\;,
\quad\quad
z_\infty=\frac{r_+ - r'_-}{r_+ - r_-}.
$$
Again we can factor out the singularity at $z=z_\infty$ by the 
transformation as 
\begin{equation}
R(z)=z^{B_1} (z-1)^{B_2} (z-z_r)^{B_3} (z-z_\infty)^{2s+1} g(z),
\end{equation}
where 
\begin{eqnarray}
B_1 &=& \frac{1}{2}\left\{-s \pm i\left[
\frac{2(1+\alpha)a^2
\left(\omega(r_-^2 + a^2)-am-\frac{eQr_-}{1+\alpha}\right)}
{\alpha(r'_+ - r_-)(r'_- - r_-)(r_+ - r_-)}
- is \right]\right\}, \nonumber \\
B_2 &=& \frac{1}{2}\left\{-s \pm i\left[
\frac{2(1+\alpha)a^2
\left(\omega(r_+^2 + a^2)-am-\frac{eQr_+}{1+\alpha}\right)}
{\alpha(r'_+ - r_+)(r'_- - r_+)(r_- - r_+)}
- is \right]\right\}, \nonumber \\
B_3 &=& \frac{1}{2}\left\{-s \pm i\left[
\frac{2(1+\alpha)a^2
\left(\omega({r'_+}^2 + a^2)-am-\frac{eQr'_+}
{1+\alpha}\right)}
{\alpha(r_- - r'_+)(r'_- - r'_+)(r_+ - r'_+)}
- is \right]\right\}. \nonumber 
\end{eqnarray}
Then, $g(z)$ satisfies the Heun's  equation as
\begin{eqnarray}
&&\makebox[-2cm]{}
\left\{ \ \frac{d^2}{dz^2} + \left[\frac{2B_1+s+1}{z} + \frac{2B_2+s+1}{z-1}
+\frac{2B_3+s+1}{z-z_r} \right] \frac{d}{dz} \right. \nonumber \\ && \makebox[5cm]{}\left.
+ \frac{\sigma_+ \sigma_- z + v}{z (z-1)(z-z_r)} \ \right\} g(z) =0.
\label{eqn:HR}
\end{eqnarray}
where 
\begin{eqnarray}
\sigma_\pm \makebox[-3mm]{}&=&\makebox[-3mm]{} B_1 + B_2 + B_3 + 2s+1 
+\frac{1}{2}\left\{-s \pm i\left[
\frac{2(1+\alpha)a^2
\left(\omega({r'_-}^2 + a^2)-am-\frac{eQr'_-}
{1+\alpha}\right)}
{\alpha(r_+ - r'_-)(r_- - r'_-)(r'_+ - r'_-)}
- is \right]\right\}, \nonumber \\
v &=&   \frac{2a^4 (1+\alpha)^2}{\alpha^2 {\cal D}}
        \frac{(r_+ - r'_+)^2 (r_+ - r'_-)^2 (r_- - r'_-) (r'_+ - r'_-)}
{r_+ - r_-} \nonumber \\
&& \makebox[-7mm]{}
\Bigg\{-\omega^2 r_-^3(r_+ r_- - 2r_+ r'_+ + r_- r'_+)
+2a\omega(a\omega-m)r_- (r_+ r'_+ - r_-^2) \nonumber \\
&&  \makebox[7cm]{}-a^2(a\omega-m)^2(2r_- - r_+ - r'_+) \nonumber \\
&& \makebox[-7mm]{}
+\frac{eQ}{1+\alpha} 
\big[ \ \omega{r_-}^2(r_+ r_- + r_-^2 -3r_+ r'_+ + r_- r'_+)
- a(a\omega-m)(r_+ r_- -3r_-^2 +r_+ r'_+ + r_- r'_+)
\  \big]  \nonumber \\
&& + \left(\frac{eQ}{1+\alpha}\right)^2 r_-(-r^2_- +r_+ r'_+) \ \Bigg\} 
\nonumber \\
&& \makebox[-7mm]{}
+\frac{2isa^2(1+\alpha)}{\alpha}
\frac{\left[ \omega (r_- r'_- +a^2)-am 
-\frac{eQ}{1+\alpha}\frac{r_- +r'_-}{2} \right]}
{(r_+ - r_-)(r'_+ - r'_-)(r_- - r'_-)}  \nonumber \\
&& \makebox[-5mm]{}
+(s+1)(2s+1)\left[\frac{2{r'_-}^2}
{(r_+ - r_-)(r'_+ -r'_-)}-z_\infty\right]
\nonumber \\
&& \makebox[-3mm]{}
- 2B_1(z_r B_2+B_3)-(s+1)\left[ (1+z_r)B_1+z_rB_2+B_3 \right] \nonumber \\
&& -\frac{a^2}{\alpha(r_+ - r_-)(r'_+ - r'_-)} 
\left[-\lambda-2iseQ+2s(1-\alpha)\right].
\end{eqnarray}
Here ${\cal D}$ is the discriminant of $\Delta_r = 0$,
\begin{eqnarray}
{\cal D}&=&(r_+ - r_-)^2 (r_+ - r'_+)^2 (r_+ - r'_-)^2
(r_- - r'_+)^2 (r_- - r'_-)^2 (r'_+ - r'_-)^2 \nonumber \\
&=& \frac{16a^{10}}{\alpha^5}\Bigg\{ \ 
(1-\alpha)^3\left[M^2-(1-\alpha)(a^2+Q^2)\right] \nonumber \\ &&  +\frac{\alpha}{a^2}\left[-27M^4+36(1-\alpha)M^2(a^2+Q^2)
- 8(1-\alpha)^2(a^2+Q^2)^2\right]
-\frac{16\alpha^2}{a^4}(a^2+Q^2)^3 \ \Bigg\}. \nonumber
\end{eqnarray}

The sign ambiguity in $B_2$ or $B_3$ are related to the boundary 
condition at the horizon or at the de Sitter horizon, respectively. 
We can take either one of signs of $B_1$. 

\section{Solution of Teukolsky equation}
\setcounter{equation}{0}

In the former section, we have shown that both the angular and radial 
Teukolsky equations can be transformed into Heun's equations 
by factoring out the singularity at $z_\infty$.
Thus, both equations can be solved in a similar way by taking into 
account the boundary conditions.
First we briefly explain the formal prescription solving the Heun's 
equation by a series expansion of the hypergeometric functions  
and then consider each case including the boundary conditions.


\subsection{Solution of Heun's equation} \label{sec:sol}
The standard form of the Heun's equation with regular singularities 
at 0, 1, $a_H$ and $\infty$ is given by[12]
\begin{equation}
\left\{ \frac{d^2}{dz^2} + \left[\frac{\gamma}{z} + \frac{\delta}{z-1}
+\frac{\epsilon}{z-a_H} \right] \frac{d}{dz}
+ \frac{\alpha\beta z - q}{z (z-1)(z-a_H)}  \right\} y_\nu(z) =0,
\label{eqn:HE}
\end{equation}
where parameters should satisfy the condition, 
$\alpha+\beta+1=\gamma+\delta+\epsilon$ and we assume $|a_H|>1$.  
Solutions which are analytic in some domain including two 
singularities are called Heun's functions. 
We particularly explain a formal construction of a Heun's function 
relative to the points 0, 1 by means of a series of hypergeometric functions. 

We first construct a solution valid in the neighborhood of 0. 
The solution is analytic in the interior of the ellipse with foci 
at 0, 1  
if the condition for augmented convergence presented bellow is satisfied.
To the end we write $y_{\nu}(z)$ in the following form,
\begin{eqnarray}
y_{\nu}(z)&=&\sum_{n=-\infty}^{+\infty} c_n^{\nu}u_{\nu+n}(z), 
\label{eqn:ynu}\\
u_{\nu}(z)&=&F(-\nu,\nu+\omega;\gamma;z)\;, \label{eqn:unu}
\end{eqnarray}
where $\omega=\gamma+\delta-1=\alpha+\beta-\epsilon$. $\omega$ is 
usually used for the definition of Heun's equation and should not 
be confused with the frequency which we use in this paper 
frequently. 

By using the following recurrence relations[12],
\begin{eqnarray}
z u_\nu(z) &=& -\frac{(\nu+\omega)(\nu+\gamma)}
{(2\nu+\omega)(2\nu+\omega+1)} u_{\nu+1}(z)
+\frac{2\nu(\nu+\omega)+\gamma(\omega-1)}
{(2\nu+\omega+1)(2\nu+\omega-1)} u_\nu(z) \nonumber \\
& & -\frac{\nu(\nu+\delta-1)}
{(2\nu+\omega)(2\nu+\omega-1)} u_{\nu-1}(z), \\
z(z-1)\frac{d}{dz} u_\nu(z) &=&  -\frac{\nu(\nu+\omega)(\nu+\gamma)}
{(2\nu+\omega)(2\nu+\omega+1)} u_{\nu+1}(z)
+\frac{\nu(\nu+\omega)(\gamma-\delta)}
{(2\nu+\omega+1)(2\nu+\omega-1)} u_\nu(z) \nonumber \\
& & -\frac{\nu(\nu+\omega)(\nu+\delta-1)}
{(2\nu+\omega)(2\nu+\omega-1)} u_{\nu-1}(z),
\end{eqnarray}
it is shown that $y_\nu(z)$ becomes a solution if 
coefficients $c_n$'s satisfy 
the following three term recurrence relations,
\begin{equation}
\alpha^\nu_n c^\nu_{n+1} + \beta^\nu_n c^\nu_n + \gamma^\nu_n c^\nu_{n-1}=0,
\label{eqn:rec}
\end{equation}
where
\begin{eqnarray}
\alpha^\nu_n &=& -\frac{(\nu+n+1)(\nu+n+\omega-\alpha+1)
(\nu+n+\omega-\beta+1)(\nu+n+\delta)}
{(2\nu+2n+\omega+2)(2\nu+2n+\omega+1)}, \label{eqn:alp} \\
\beta^\nu_n &=& 
\frac{J^\nu_n}{(2\nu+2n+\omega+1)(2\nu+2n+\omega-1)} 
- a_H(\nu+n)(\nu+n+\omega)-q, \\
\gamma^\nu_n &=& -\frac{(\nu+n+\alpha-1)(\nu+n+\beta-1)
(\nu+n+\gamma-1)(\nu+n+\omega-1)}
{(2\nu+2n+\omega-2)(2\nu+2n+\omega-1)},  \label{eqn:gam}
\end{eqnarray}
and
\begin{eqnarray}
J^\nu_n &=& \epsilon(\nu+n)(\nu+n+\omega)(\gamma-\delta) \nonumber \\
        & &   +[ \ (\nu+n)(\nu+n+\omega)+\alpha\beta \ ]
                [ \ 2(\nu+n)(\nu+n+\omega)+\gamma(\omega-1) \ ].
\end{eqnarray}
By defining the continued fractions
\begin{equation}
R_n(\nu)=\frac{c^\nu_n}{c^\nu_{n-1}}, \makebox[2cm]{}
L_n(\nu)=\frac{c^\nu_n}{c^\nu_{n+1}},
\end{equation}
they satisfy
\begin{equation}
R_n(\nu)=-\frac{\gamma^\nu_n}{\beta^\nu_n+\alpha^\nu_n R_{n+1}(\nu)}
 \;, \qquad
L_n(\nu) = -\frac{\alpha^\nu_n}{\beta^\nu_n+\gamma^\nu_n L_{n-1}(\nu)}
\;.
\end{equation}
Now we can evaluate the coefficients $c_n^{\nu}$ by using either 
series $R_n(\nu)$ or $L_n(\nu)$ with appropriate initial data. 
The convergence of the series requires a following 
transcendental equation,
\begin{equation}
R_n(\nu) L_{n-1}(\nu) = 1. \label{eqn:aug}
\end{equation}
This equation determines either the eigenvalue $\lambda$ for the angular 
equation, where the characteristic exponent $\nu$ (the renormalized angular 
momentum) can be taken zero, or $\nu$ for the radial cases, respectively. 

It will be worthwhile to note that  if $\nu$ is a solution 
of eq.(\ref{eqn:aug}), then $-\nu-\omega$ is also the solution. This 
can be shown as followings: First, we notice the algebraic relations,   
$\alpha^{-\nu-\omega}_{-n}=\gamma^\nu_n$, 
$\beta^{-\nu-\omega}_{-n}=\beta^\nu_n$ and 
$\gamma^{-\nu-\omega}_{-n}=\alpha^\nu_n$. From these relations, 
we find that  $c^{-\nu-\omega}_{-n}$'s 
satisfy the same recurrence relations as these for $c^\nu_n$'s. 
If we choose $c^\nu_0=c^{-\nu-\omega}_0$, we have 
$c^\nu_n=c^{-\nu-\omega}_{-n}$ so that 
\begin{equation}
R_n(-\nu-\omega) L_{n-1}(-\nu-\omega)=
R_n(\nu) L_{n-1}(\nu)= 1.
\end{equation}
From eq.(4.3), we see that $y_{-\nu-\omega}(z)$ agrees with $y_{\nu}(z)$ 
by choosing $c^\nu_0=c^{-\nu-\omega}_0$. 
 

\subsection{Solution of the angular equation} 
By comparing Heun's equation (4.1) with eq.(3.4), parameters to 
define Heun's equation, $\alpha, \beta, \gamma, 
\delta, \epsilon, \omega,  q$ and $a_H$ are given by
\begin{eqnarray}
\label{eqn:Sexp}
\begin{array}{lll}
\alpha=\rho_+\;,  & \beta=\rho_-\;,&\gamma=2A_1+1\;,   \\
 \delta = 2A_2+1\;, & \epsilon = 2A_3+1\;, &\omega=2(A_1+A_2)+1\;,\\
q= -u\;, & a_H = z_s\;,&  
\end{array}
\end{eqnarray}
We require that $f(z)$ is a regular function at $z=0$, 1, 
which is satisfied by taking $\nu=0$. 
Then the solution  is given in the form of 
the series of the Jacobi polynomials extending from $n=0$ to $\infty$, 
similarly to the Kerr geometry case, 
\begin{eqnarray} f(z) \makebox[-3mm]{}&=&\makebox[-3mm]{} 
\sum_{n=0}^{\infty} a_n u_n(z), \\ u_n(z) \makebox[-3mm]{}
&=&\makebox[-3mm]{} F(-n,n+\omega;\gamma;z) 
= (-)^n \frac{\Gamma(2n+\omega)n!}{\Gamma(n+\gamma)} 
P^{(\omega-\gamma,\gamma-1)}_n (2z-1).
\end{eqnarray}
The coefficients $a_n$'s are determined by solving the equation 
for $L_n(\nu)$ in eq.(\ref{eqn:rec})  with $L_{-1}(\nu)=0$, where 
$a_0$ gives the overall normalization. 
The eigenvalue $\lambda$ can be obtained by transcendental equation 
(\ref{eqn:aug}) with the condition $R_{\infty}^{\nu}=L_{-1}^{\nu}=0$. 
We give $\lambda$ up to $O(\xi^3)$ and the coefficient of each power of $\xi$ 
up to $O(\alpha^2)$ 
as 
\begin{eqnarray}
\lambda &=&
l(l+1)-s^2+s \nonumber \\
&& +\alpha\left[ \ -l(l+1)+s^2-s+2m^2+
\frac{2m^2s^2}{l(l+1)}-l^2 H(l)+(l+1)^2H(l+1) \ \right] \nonumber \\ && 
+\left\{-2m\left(1+\frac{s^2}{l(l+1)}\right) \right. \nonumber \\ && \left.
- 2m\alpha\left[1+\frac{s^2}{l(l+1)}
-\left(1+\frac{s^2}{(l-1)(l+1)}\right)H(l)
+\left(1+\frac{s^2}{l(l+2)}\right)H(l+1)\right] 
\right\} \xi \nonumber \\ && +\Bigg\{ H(l+1)-H(l) 
+\alpha\Bigg[H(l+1)-H(l)+2\Big((l+1)^2H(l+1)-l^2H(l)\Big) 
\nonumber \\
&& \makebox[5mm]{}  
- lH^2(l)+(l+1)H^2(l+1)-\frac{H(l)H(l+1)}{l(l+1)} \nonumber \\
&& \makebox[5mm]{}
+6m^2s^2\Big(\frac{H(l+1)}{l(l+1)^2(l+2)} 
-\frac{H(l)}{(l-1)l^2(l+1)}\Big) \nonumber \\
&& \makebox[5mm]{}
+4m^2s^4\Big(\frac{H(l+1)}{l^2(l+1)^2(l+2)^2}
-\frac{H(l)}{(l-1)^2l^2(l+1)^2}\Big)
\Bigg]\Bigg\}\xi^2, 
\end{eqnarray}
where 
\begin{eqnarray}
H(l) \makebox[-3mm]{}&=&\makebox[-3mm]{} 
\frac{2(l^2-m^2)(l^2-s^2)^2}{(2l-1)l^3(2l+1)},
\end{eqnarray}
and $l$ is the angular momentum which takes an integer or half integer 
number satisfying $l \geq \max(|m|,|s|)$. 
From the result it seems that $\lambda-s(1-\alpha)$ is 
an even function of $s$.
Indeed this holds true because eq.(\ref{eqn:aug}) which determines 
$\lambda$ contains $\alpha_n, \beta_n$ and $\gamma_n$ in the forms of 
$\alpha_{n-1} \gamma_n$ and 
$\beta_n+\frac{i}{4\sqrt{\alpha}}(\lambda-s(1-\alpha))$,  
which are invariant under $s \rightarrow -s$, respectively.

In the rest of this section, we consider two limits, the Kerr-Newman limit  
and the Reissner-Nordstr\"om-de Sitter limit. It may be worthwhile to 
note that the angular Teukolsky equation (3.1) does not contain 
mass and charge parameters of the black hole so that the above limits 
are only meaningful ones. 

\noindent
{\it The Kerr-Newman limit}: $\Lambda=(3/a^2)\alpha 
\rightarrow 0$

In the limit, parameters behave as
\begin{eqnarray}
\begin{array}{lll}
\rho_+ = A_1+A_2 \pm s+1+O(\sqrt{\alpha}),& 
\rho_- = \displaystyle{\pm \frac{i}{\sqrt{\alpha}} \xi +O(1)},& 
A_3 = \displaystyle{\pm \frac{i}{2\sqrt{\alpha}}\xi +O(1)}, \\
\multicolumn{3}{l}{
u =  \displaystyle{-\frac{i}{4\sqrt{\alpha}} \left[ 
\lambda-2A_1 A_2 -A_1 -A_2 +2\left(m+s \mp (2A_1 +1)\right)\xi 
-\frac{m^2-s^2}{2}-s \right]},} \\
z_s = \displaystyle{-\frac{i}{4\sqrt{\alpha}} +O(1)},& 
z_\infty = \displaystyle{-\frac{i}{2\sqrt{\alpha}} +O(1)},& \
\end{array}
\end{eqnarray}
then parameters in the recurrence relation (\ref{eqn:alp})
$-$(\ref{eqn:gam}) become by omitting the superscript $\nu$ as 
\begin{eqnarray}
\alpha_n &=&
\pm \frac{i}{\sqrt{\alpha}}\xi \ 
\frac{(n+1)(n+A_1+A_2 \mp s +1)(n+2A_2+1)}
{2(2n+2A_1+2A_2+3)(n+A_1+A_2+1)}, \\
\beta_n &=& 
\frac{i}{\sqrt{\alpha}} \left\{
\pm \xi \frac{j_n}
{2(n+A_1+A_2+1)(n+A_1+A_2)} \right.\nonumber \\
&& \makebox[10mm]{} + \frac{n(n+2A_1+2A_2+1)}{4} \\
&& \makebox[10mm]{} \left. -\frac{1}{4}\left[ 
\lambda-2A_1 A_2 -A_1 -A_2 +2\big(m+s \mp (2A_1 +1)\big)\xi 
-\frac{m^2-s^2}{2}-s \right] \right\}, \nonumber \\
\gamma_n &=& 
\mp\frac{i}{\sqrt{\alpha}}\xi \ 
\frac{(n+A_1+A_2\pm s)(n+2A_1)(n+2A_1+2A_2)}
{2(2n+2A_1+2A_2-1)(n+A_1+A_2)},
\end{eqnarray}
where
\begin{eqnarray}
j_n &=& n(n+2A_1+2A_2+1)(A_1-A_2) \nonumber \\
& & +(A_1+A_2 \pm s+1)
[ \ n(n+2A_1+2A_2+1)+(2A_1+1)(A_1+A_2) \ ] \nonumber
\end{eqnarray}
Thus, the recurrence relations for $a_n$'s are obtained from 
those for $c_n$'s in eq.(\ref{eqn:rec}) by dropping the infinite 
factor $\displaystyle{\frac{i}{\sqrt{\alpha}}}$. This recurrence 
relation agrees  with those in Ref.4 for the Kerr geometries. 
In this limit, the Heun's equation (\ref{eqn:HS}) reduces to a 
confluent Heun's equation which has two regular singularities 
at $z=0, 1$ and an irregular singularity at $\infty$. 
The angular wave function $S(z)$ is given by using $f(z)$ as
\begin{equation}
S(z) \longrightarrow 
\left(-\frac{i}{4\sqrt{\alpha}}\right)^{\frac{i}{2\sqrt{\alpha}} \xi} 
z^{A_1} (z-1)^{A_2} e^{\mp 2\xi z} f(z)\;.
\end{equation}

\noindent 
{\it The Reissner-Nordstr\"om-de Sitter limit}: $a \rightarrow 0$ 

In this limit, we have to take $\alpha \rightarrow 0$ simultaneously, 
in order to keep $\Lambda=3\alpha/a^2$ fixed. 
Then, parameters become the following forms;
\begin{eqnarray}
\begin{array}{lll}
\rho_+ = A_1+A_2\pm s+1\;,& 
\rho_- = \displaystyle{A_1+A_2 
\pm \sqrt{\frac{3}{\Lambda}}\omega+1}\;,& 
A_3 = \displaystyle{\pm\frac{i}{2}}\sqrt{\frac{3}{\Lambda}}\omega\;, \\
\multicolumn{3}{l}{\displaystyle{
u=  \frac{-i}{4a}\sqrt{\frac{3}{\Lambda}} \left(
\lambda-2A_1A_2-A_1-A_2-\frac{m^2-s^2}{2}-s \right)\;,}} \\
z_s = \displaystyle{\frac{-i}{4a}\sqrt{\frac{3}{\Lambda}}}\;,& 
z_\infty = \displaystyle{\frac{-i}{2a}\sqrt{\frac{3}{\Lambda}}}\;.& 
\end{array}
\end{eqnarray}
Then, $\alpha^\nu_n$ and $\gamma^\nu_n$ are $O(1)$, while 
$\beta^\nu_n$ are $O(1/a)\to \infty$,  
\begin{equation}
\beta^\nu_n = \frac{i}{4a}\sqrt{\frac{3}{\Lambda}}\left[
n(n+2A_1+2A_2+1)-\lambda+2A_1A_2+A_1+A_2 +\frac{m^2-s^2}{2}+s \right].
\end{equation}
In order that coefficients $a_n$ have definite limits, 
$\beta^\nu_n$ must be 0 for a specific value of $n$.
The eigenvalue $\lambda$ is determined by this condition as
\begin{equation}
\lambda=(l-s+1)(l+s), \label{eqn:lambda-ds}
\end{equation}
where $l=n+A_1+A_2 \geq \max(|m|,|s|)$.
Therefore the angular wave function has the following form,
\begin{eqnarray}
S(z)&\rightarrow&\frac{i}{2a}\sqrt{\frac{3}{\Lambda}} 
\left( \frac{i}{4a}\sqrt{\frac{3}{\Lambda}} \right)^
{\pm\frac{i}{2}\sqrt{\frac{3}{\Lambda}}\omega} z^{A_1} (z-1)^{A_2}\\
 && \times F(-l+A_1+A_2, l+A_1+A_2+1; 2A_1+1; z)\;. 
\end{eqnarray}
In this limit, the Heun's equation (\ref{eqn:HS}) reduces to a 
hypergeometric equation. These results coincide with that of [14].

In the former subsection, we comment that if $\nu$ is a solution 
of eq.(4.13), then $-\nu-\omega$ is a solution. For the angular 
solution, we chose $\nu=0$ and we know that $\omega$=integer so 
that this transformation of the exponent does not lead any new 
solution. 

\subsection{Solution of the radial equation} 

By comparing eq.(\ref{eqn:HE}) with eq.(\ref{eqn:HR}), we find 
parameters in Heun's equation, $\alpha, \beta, \gamma, 
\delta, \epsilon, q$ and $a_H$ is given by
\begin{eqnarray}
\label{eqn:Rexp}
\begin{array}{lll}
\alpha=\sigma_+,  & \beta=\sigma_-,  & \\
\gamma=2B_1+s+1, & \delta = 2B_2+s+1, & \epsilon = 2B_3+s+1, \\
q= -v, & a_H = z_r. &  
\end{array}
\end{eqnarray}
The solution convergent in the ellipse with foci at $z=0, 1$ which 
 correspond to 
$r=r_-, r_+$ respectively is given by 
\begin{eqnarray} g_\nu(z)&=&\sum_{n=-\infty}^{+\infty} b^\nu_n u_{\nu+n}(z), 
\label{eqn:gnu}\\ u_\nu(z)&=&F(-\nu,\nu+2(B_1+B_2+s)+1;2B_1+s+1;z).
\end{eqnarray}
where $\omega=\gamma+\delta-1=2(B_1+B_2+s)+1$. 
The expansion coefficients $b^\nu_n$'s are determined by the recurrence 
relation (\ref{eqn:rec}) in which $c_n$'s are replaced with $b^\nu_n$'s.
The radial solution is expressed by a series where $n$ runs from $-\infty$ 
to $+\infty$. 
This is because the eigenvalue $\lambda$ is fixed from 
the angular solution and thus the characteristic exponent 
(the renormalized momentum) $\nu$ is determined by the 
transcendental equation (\ref{eqn:aug}) which guarantees the 
convergence of series. Remind that if $g_{\nu}(z)$ is a solution 
then $g_{-\nu-\omega}(z)$ is a solution too. 

\noindent
{\it  the Kerr-Newman limit: $\Lambda=(3/a^2)\alpha 
\rightarrow 0$}

In this limit, $r_\pm$ and $r'_\pm$ are
\begin{eqnarray}
r_\pm &\longrightarrow& r^0_\pm \left[ 1+\frac{r^0_\pm ({r^0_\pm}^2+a^2)}
{2a^2(r^0_\pm-M)} \alpha \right]
+ O(\alpha^2), \\
r'_\pm &\longrightarrow& \pm \frac{a}{\sqrt{\alpha}} \left[
1\pm \frac{M}{a} \sqrt{\alpha} +\frac{Q^2-3M^2}{2a^2} \alpha \right] 
+O(\alpha),
\end{eqnarray}
then parameters are
\begin{eqnarray}
B_1 &=& 
\frac{1}{2}\left[
- s\mp i(-\tilde{\epsilon}+\tilde{\tau}+is)\right]+O(\sqrt{\alpha})
\equiv B^0_1+O(\sqrt{\alpha}), \nonumber \\
B_2 &=& \frac{1}{2}\left[
- s\pm i(\tilde{\epsilon}+\tilde{\tau}-is)\right]+O(\sqrt{\alpha})
\equiv B^0_2+O(\sqrt{\alpha}), \nonumber \\
B_3 &=&
\mp  \displaystyle{\frac{ia\omega}{2\sqrt{\alpha}}
+\frac{1}{2}\left[-s\pm
i(-\tilde{\epsilon}-is)\right]+O(\sqrt{\alpha})}, \nonumber \\
\sigma_+ &=&
B^0_1+B^0_2+s +1\pm i[-\tilde{\epsilon}-is]+O(\alpha), \nonumber \\
\sigma_- &=&
\mp \frac{ia\omega}{\sqrt{\alpha}}
+B^0_1+B^0_2+s+1+O(\sqrt{\alpha}),  \\
v &=& -\frac{a}{2\sqrt{\alpha}(r^0_+ -r^0_-)} \left\{
-\lambda+2s-2iseQ+2B^0_1B^0_2+(s+1)\left(B^0_1+B^0_2\right) 
\right.\nonumber \\ & & \left. -i\epsilon \tilde{\kappa}
\left[s\pm (2B^0_1+s+1)\right]
+2\epsilon^2-\epsilon^2\tilde{\kappa}+eQ\epsilon(\tilde{\kappa}-3)
-\epsilon\tilde{q}
-\frac{1}{2}\left( \tilde{\epsilon}^2+\tilde{\tau}^2
- 2is\tilde{\epsilon} \right)
\right\} \nonumber \\
& & +O(\sqrt{\alpha}),  \nonumber \\
z_r &=& \frac{a}{2\sqrt{\alpha}(r^0_+ -r^0_-)}+O(1),\makebox[2cm]{} 
z_\infty= \displaystyle{\frac{a}{\sqrt{\alpha}(r^0_+ -r^0_-)}+O(1)}, 
\end{eqnarray}
where $\tilde{\epsilon}=2M\omega-eQ=\epsilon-eQ$, $\epsilon=2M\omega$, 
$\tilde{\kappa}=\sqrt{1-\frac{a^2+Q^2}{M^2}}$, 
$\tilde{q}=\frac{am+Q^2\omega}{M}$ and 
$\tilde{\tau}=\displaystyle{
\frac{\tilde{\epsilon}-\tilde{q}}{\tilde{\kappa}}}$. 
The parameter $\omega$ in the above expressions is used for 
the frequency. 
The singularity at $\infty$ becomes irregular because of 
$z_r \rightarrow \infty$. Thus the radial equation (\ref{eqn:HR}) 
becomes to a confluent Heun's equation as the angular one does. 
In the Kerr limit ($\Lambda \to \infty$ and $Q=0$), coefficients of 
the recurrence relation (\ref{eqn:rec}) coincide with those in Ref.6, 
if we divide them 
by $\displaystyle{-\frac{a}{2\sqrt{\alpha}(r^0_+ -r^0_-)}}$.

\noindent
{\it the de Sitter limit: $M, Q$ and $a \rightarrow 0$}

We rearrange the locations of the singularities so that 
$z=0, 1, z'_r,$ and $\infty$ correspond to $r=r'_+, r_-, r_+$ and 
$r'_-$ respectively;
\begin{equation}
z=\frac{(r_- -r'_-)}{(r_- -r'_+)}\frac{(r -r'_+)}{(r-r'_-)},
\end{equation}
thus $\displaystyle{z'_r=\frac{(r_- -r'_-)}{(r_- -r'_+)}
\frac{(r_+ -r'_+)}{(r_+ -r'_-)}}$.
The effect of this rearrangement is only to replace $r_-, r_+, r'_+$ 
and $r'_-$ with $r'_+, r_-, r_+$ and $r'_-$, respectively, 
in the radial equation.
In this limit we find 
\begin{equation} 
r_\pm \rightarrow \pm ia, \hspace{1cm} 
r'_\pm \rightarrow \pm \sqrt{\frac{3}{\Lambda}}, \hspace{1cm}
\end{equation}
and 
\begin{eqnarray}
B_1 &=& \frac{1}{2}      
\left[-s \mp
i\left(\sqrt{\frac{3}{\Lambda}}\omega+is\right)\right],
\nonumber  \\
B_2 &=& \frac{1}{2}\left[-s \pm (m+s)\right], \hspace{1cm}
B_3=\frac{1}{2}\left[-s \pm (-m+s)\right], \nonumber \\
       \sigma_\pm &=& B_1+B_2+B_3+2s+1
          +\frac{1}{2}\left[-s \pm i 
              \left(\sqrt{\frac{3}{\Lambda}}\omega-is\right)\right],  \\ 
v &=& -\lambda+2s-(s+1)(2s+1)-2B_1(B_2+B_3)-(s+1)(2B_1+B_2+B_3)
         +is\sqrt{\frac{3}{\Lambda}}\omega, \nonumber \\ z'_r &=& 1,  
\hspace{1cm}  z_\infty=-1. \nonumber \end{eqnarray}
The resulting radial equation is the following 
with the three singularities,
\begin{equation}
\left\{\frac{d^2}{dz^2}
+\left[\frac{2B_1+s+1}{z}+\frac{2(B_1+B_2+s+1)}{z-1}\right]
\frac{d}{dz} +\frac{\sigma_+\sigma_- +v}{z(z-1)^2}\right\}g(z)=0.
\end{equation}
By using $\lambda=(l-s+1)(l+s)$ (\ref{eqn:lambda-ds}) and setting  
$\displaystyle{g(z)=(z-1)^{-B_2-B_3-s-\frac{1}{2}+(l+\frac{1}{2})} 
g_{\mbox{{\scriptsize deS}}}(z)}$, 
this equation reduces to a hypergeometric equation thus we obtain  
\begin{eqnarray}
R(z) \makebox[-2mm]{}&=&\makebox[-2mm]{}
z^{B_1} (z-1)^{l-s} (z+1)^{2s+1} g_{\mbox{{\scriptsize deS}}}(z) 
\nonumber \\
\makebox[-2mm]{}&=&\makebox[-2mm]{}
z^{B_1} (z-1)^{l-s} (z+1)^{2s+1} 
F(l+1-s,l+1+i\sqrt{\frac{3}{\Lambda}}\omega;
1-s+i\sqrt{\frac{3}{\Lambda}}\omega;z).
\end{eqnarray}
This radial function coincides with that in Ref.13 except for 
$\omega \rightarrow -\omega$ because of a difference between 
our choice of the null tetrad and theirs.


\section{Conclusions and discussions}

In this note, we have shown that equations for perturbations of 
the Kerr-de Sitter black hole for massless particles with any spins 
are reduced to the Heun's equation and thus we can obtain solutions 
in the form of series of hypergeometric functions. This method 
has been extended to the cases of Kerr-Newman-de Sitter geometries 
except for photons and gravitons for which the technique to 
separate angular and radial parts are not succeeded so far.  
We have explicitly constructed solutions for angular and radial 
equations and examined some limiting cases. We have shown that these 
limits reproduce the former known results.

The radial solutions we obtained in the text are valid 
inside a ellipse with foci at 
$z=0$, 1 and are not valid at de Sitter horizon.  In order 
to apply for some physical situations, we have to construct 
solutions which are valid in the entire regions of $r$ and 
satisfy some specific boundary condition. For this, we 
should obtain solutions valid around the de Sitter horizon and 
match solutions with different convergence regions in the region 
where both solutions are convergent. This procedure was made 
for the Kerr black hole case in Ref.6,7. Also, it may be interesting 
to see the Teukolsky-Starobinsky identities analytically. 
We will answer these questions in the future publication. 


\end{document}